\documentclass[twocol]{ametsocV6.1}

\usepackage{caption, subcaption, float}
\usepackage{subeqnarray}
\usepackage{gensymb}
\usepackage{multirow}
\allowdisplaybreaks
\usepackage{color}
\usepackage{booktabs}

\usepackage{dirtytalk}

\newcommand{\derp}[2]{\frac{\partial #1}{\partial #2}}
\newcommand{\der}[2]{\frac{d #1}{d #2}}

\newcommand{\bse}{\begin{subeqnarray}}
\newcommand{\ese}{\end{subeqnarray}}

\newcommand{\e}[1]{\cdot 10 ^{#1}}

\title{Physics of an AMOC Overshoot in a Box Model}

\authors{Jelle Soons,\aff{a}\correspondingauthor{Jelle Soons j.soons@uu.nl} Tobias Grafke, \aff{b} Ren\'e M. van Westen, \aff{a}
and Henk A. Dijkstra\aff{a} 
}

\affiliation{\aff{a}{Institute for Marine and Atmospheric Research, Utrecht University, Princetonplein 5, 3584 CC Utrecht, The Netherlands}\\
\aff{b}{Mathematics Institute, University of Warwick, Coventry CV4 7AL, United Kingdom}
}

\abstract{Recently the global average temperature has temporarily exceeded the 1.5$^\circ$C goal of the Paris Agreement, and so an overshoot of various climate tipping elements becomes increasingly likely. In this study we analyze the physical processes of an overshoot of the Atlantic Meridional Overturning Circulation (AMOC), one of the major tipping elements, using  a conceptual box model.  Here either the atmospheric temperature above the North Atlantic, or the freshwater forcing into the North Atlantic overshoot their respective critical boundaries. In both cases a higher forcing rate can prevent a collapse of the AMOC, since a higher rate of forcing causes initially a fresher North Atlantic, which in turn results in a higher northward transport by the subtropical gyre supplementing the salinity loss in time. For small exceedance amplitudes the AMOC is still resilient as the forcing rates can be low and so other state variables outside of the North Atlantic can adjust. Contrarily, for larger overshoots the trajectories are dynamically similar and we find a lower limit in volume and exceedance time for respectively freshwater and temperature forcing in order to prevent a collapse. Moreover, for a large overshoot an increased air-sea temperature coupling has a destabilizing effect, while the reverse holds for an overshoot close to the tipping point. The understanding of the physics of the AMOC overshoot behavior is important for interpreting results of Earth System Models and for evaluating the effects of mitigation and intervention strategies.}
\begin{document}

\maketitle

%
%
%
%
%

%
\section{Introduction}
The Atlantic Meridional Overturning Circulation (AMOC) is of vital importance in regulating the Earth's climate. It consists of a northward flow of relatively warm upper ocean water, and a deeper southward transport of relatively cool water within the Atlantic basin. This northward heat transport of about 1.5 PW \citep{johns2011continuous} is a major reason for the rather mild climate in Western Europe. A key factor in this overturning circulation is believed to be the water mass transformation in the subpolar North Atlantic Ocean. Here the relatively warm and salty sea water originating from low latitudes is cooled by the 
 overhead atmosphere, and forms the cold and salty North Atlantic Deep Water (NADW). This dense water mass sinks and returns southward at greater depth \citep{frajka2019atlantic}.

The AMOC is considered to be a major tipping element in the present-day climate \citep{armstrong2022exceeding}, meaning that a small change in external forcing can result in hard-to-reverse transitions of the AMOC. The salt-advection feedback is responsible for this non-linear response in AMOC strength, which can cause the AMOC to tip to a substantially weaker state \citep{marotzke2000abrupt, weijer2019stability}. When the AMOC weakens, it transports less salt northward, which will inhibit the deep-water formation there and hence decrease its strength even further. Due to this positive feedback the AMOC could switch from its current \say{on} state to an \say{off} state under sufficient freshwater hosing in the North Atlantic. In this \say{off} state the AMOC is weaker or even non-existent, which has severe impacts for the global climate \citep{bellomo2024impacts, van2024physics}.

\citet{stommel1961thermohaline} was the first to realize that the AMOC could have two stable states using a simple two-box model. For a range of the same buoyancy surface forcings both \say{on} and \say{off} state are stable. If these surface forcings are altered such that the southern (northern) box retains increasingly denser water then the \say{on} (\say{off}) state disappears in a saddle-node bifurcation. Since then, similar bifurcation structures have been found in increasingly more complex AMOC models, where multiple stable states exists under the same forcing conditions \citep{rooth1982hydrology, rahmstorf1996freshwater, dijkstra1997symmetry, lucarini2005thermohaline, dijkstra2007characterization}. For more detailed models it is no longer numerically feasible to compute these bifurcation diagrams. Then an impression of the multiple equilibrium regime is obtained by slowly increasing the freshwater forcing in the Northern Atlantic until a critical threshold is reached after which the stable AMOC state ceases to exist. This circulation will then collapse to the weaker stable state in a bifurcation-induced tipping event. Subsequently, the forcing is reduced again until the inactive AMOC state destabilizes and the original AMOC recovers. This yields a hysteresis curve where its width indicates the extent of the multiple equilibrium regime. These hysteresis curves of the AMOC strength have been found in even more elaborate models ranging from Earth system Models of Intermediate Complexity \citep{rahmstorf2005thermohaline}, to low resolution atmosphere-ocean coupled general circulation models \citep{hawkins2011bistability} and finally even in a state-of-the-art global climate model \citep{van2023asymmetry}. These results are strikingly consistent with the results provided by the conceptual models \citep{dijkstra2024role}.

A crucial task is to estimate where the boundaries of this multiple equilibrium regime are located as currently the salinity in the North Atlantic Ocean is decreasing \citep{li2021persistent}. Several studies indicate that we are currently approaching this tipping point of the AMOC based on early warning signals of historical AMOC reconstructions \citep{boers2021observation}, and based on observations of the AMOC-induced freshwater transport at the southern boundary of the Atlantic \citep{van2024physics}. Moreover, an estimate has been provided that this tipping point will be reached halfway this century \citep{ditlevsen2023warning}, but with the remark that this estimate has large uncertainties \citep{ben2024uncertainties}. 

Crossing this tipping threshold would not necessarily result in a collapse of the AMOC when the forcing would be reduced after the overshoot. If this reduction is at a sufficiently fast rate the AMOC can still be stabilized, and a \say{safe overshoot} has taken place \citep{ritchie2021overshooting, ritchie2023rate}. The possibility of a safe overshoot becomes increasingly more relevant as capabilities for negative emissions grow, such as solar radiation management and carbon capture techniques \citep{caldeira2013science, wilberforce2021progress}. Hence it becomes increasingly important to identify which physical processes determine whether an overshoot is safe.

In this paper we adopt a conceptual box model of the AMOC where not only the salt-advection feedback is represented, but also the thermal and pycnocline feedback mechanisms, which both have a stabilizing effect on the AMOC. It has a bi-equilibrium regime with a stable AMOC ON state having a northward transport, and an OFF state where this transport is non-existent. In order to study an overshoot we increase the northern freshwater flux briefly beyond the critical threshold of the ON state for various values of the atmosphere-ocean coupling parameter. A similar procedure is done with an overshoot in the northern atmospheric temperature to replicate a possible polar amplification \citep{bekryaev2010role}. This will highlight the difference between a freshwater and a temperature overshoot.

In section \ref{sec:model} we briefly introduce the box model by \citet{van2024role}. Then, section \ref{sec:fresh} discusses an AMOC overshoot where the freshwater forcing temporarily exceeds the tipping point, while section \ref{sec:temp} considers an overshoot with the northern atmospheric temperature exceeding the threshold. Section \ref{sec:sum} contains a summary and discussion of our results.

\section{Model}\label{sec:model}
\begin{figure}
    \centering
    \includegraphics[width = 19pc]{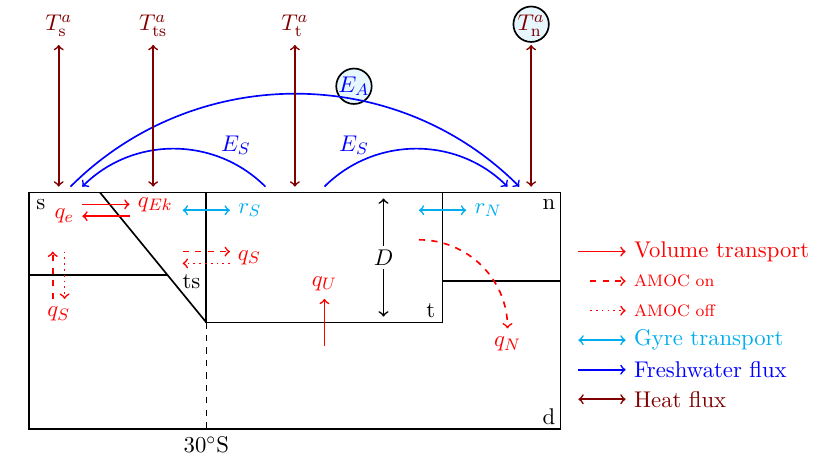}
    \caption{Sketch of the box model: the red, cyan and blue arrows represent the volume fluxes between the boxes, the transport by the subtropical gyres, and the freshwater exchange between the ocean's surface and the atmosphere, respectively. The brown arrows represent the heat exchange between the ocean's surface and the atmosphere. The boxes' labels are indicated in black. The encircled variables are the bifurcation parameters.}
    \label{fig:model}
\end{figure}

We use the box model originally introduced by \cite{cimatoribus2014meridional}, and later expanded upon by \cite{castellana2019transition},  \cite{van2024role} and by \cite{van2025saddle}. An overview of the model is presented in figure \ref{fig:model}. It represents the Atlantic Ocean sector in five boxes. A deep box (labeled \textit{d}) extends throughout the whole latitudinal width of the Atlantic basin. The upper ocean layer is separated from the rest of the ocean by a pycnocline with depth $D$. This pycnocline layer is represented by the two tropical boxes \textit{t} and \textit{ts}, where the latter is located south of $30\degree$S. This splitting allows for a distinction between fresh water transported by the AMOC and by the subtropical gyre circulation. Boxes \textit{s} and \textit{n} model the Southern and Northern Atlantic Ocean, respectively.

The ocean state in this model is completely characterized by the salinities $S_t, S_{ts}, S_n, S_s, S_d$ and temperatures $T_t, T_{ts}, T_n, T_s, T_d$ of the boxes, and by the pycnocline depth $D$. The boxes' volumes are $V_t, V_{ts}, V_n, V_s, V_d$, where only $V_n$ and $V_s$ are fixed, while the others depend on the pycnocline depth. Between these boxes three volume fluxes occur: a downwelling in the North Atlantic (and hence AMOC strength) $q_n$, the southern upwelling $q_s$ which is the difference between the northward wind-driven Ekman transport $q_{Ek}$ and the eddy-induced flow $q_e$, and the upwelling $q_u$ from the deep ocean into the pycnocline layer. The other salinity transports in the system are the wind-driven subtropical gyres, captured by coefficients $r_n$ and $r_s$, and the external freshwater flux, which is split in a symmetrical part $E_s$ and an asymmetrical part $E_a$. The latter essentially moves fresh water from the southern to the northern box, while conserving the system's total salinity. Additionally, the ocean's heat content is transported by the aforementioned volume fluxes and gyres as well. There is an atmospheric heat exchange with fixed heat exchange coefficient $\lambda^a$, through which the temperatures of the surface boxes are forced towards the prescribed atmospheric temperatures $T_t^a$, $T_{ts}^a$, $T_n^a$ and $T_s^a$. The model equations and the standard parameter values can be found in Appendix A and B. 

The bifurcation diagram of this model with either the asymmetric freshwater flux $E_a$ or the northern atmospheric temperature $T_n^a$ as bifurcation parameter show multiple stable equilibria. For low values of either parameter only the stable ON state is possible, where the AMOC strength $q_n$ is positive, corresponding to downwelling in the north and upwelling in the south. For high values, on the other hand, only the OFF state is stable. It is characterized by the absence of downwelling in the north, and the downwelling in the south completely compensated by upwelling near the tropics i.e. $q_n = 0$ and $q_s < 0$. In this state the AMOC has collapsed. In both bifurcation diagrams there is a regime where two stable equilibria exist. We are particularly interested in this regime with two stable
equilibria. The central research question of this study is
to explore how the system can exit  and re-enter this regime.

\section{Overshoot in Freshwater Flux}\label{sec:fresh}
\subsection{Motivating example}
We apply the following time-dependent symmetric freshwater forcing (with time $t\in\left(-\infty,\infty\right)$)
\begin{equation}\label{eq:forcing0}
    E_a(t) = E_a^0 + \frac{\Delta E_a + \varepsilon_a}{\cosh{(rt)}}
\end{equation}
to the equilibrium ON state at forcing level $E_a^0$. This forcing can be applied for various rates $r$, and is based on those used in previous overshooting analyses \citep{ritchie2021overshooting, ritchie2023rate}. At forcing level $E_a^0$ both an ON and OFF state exist, and it lies a distance $\Delta E_a$ away from the bifurcation point $E_a^b$ of the model. So for a freshwater forcing $E_a > E_a^b$ no stable ON state exists. Hence the forcing \eqref{eq:forcing0} has a maximum that exceeds the bifurcation point by $\varepsilon_a$: the exceedance amplitude. Note that the total additional freshwater $V_f$ that is added to the northern box is
\begin{equation}\label{eq:vf}
\begin{split}
    V_f &= \int_{-\infty}^\infty \left(E_a(t) - E_a^0\right)\, dt \\
    &= \frac{\pi\Delta E_a}{r} + \frac{\pi\varepsilon_a}{r}\,,
\end{split}
\end{equation}
where the latter term represents the additional amount of freshwater added beyond the bifurcation point.

\begin{figure}
    \centering
    \includegraphics[width = 19pc]{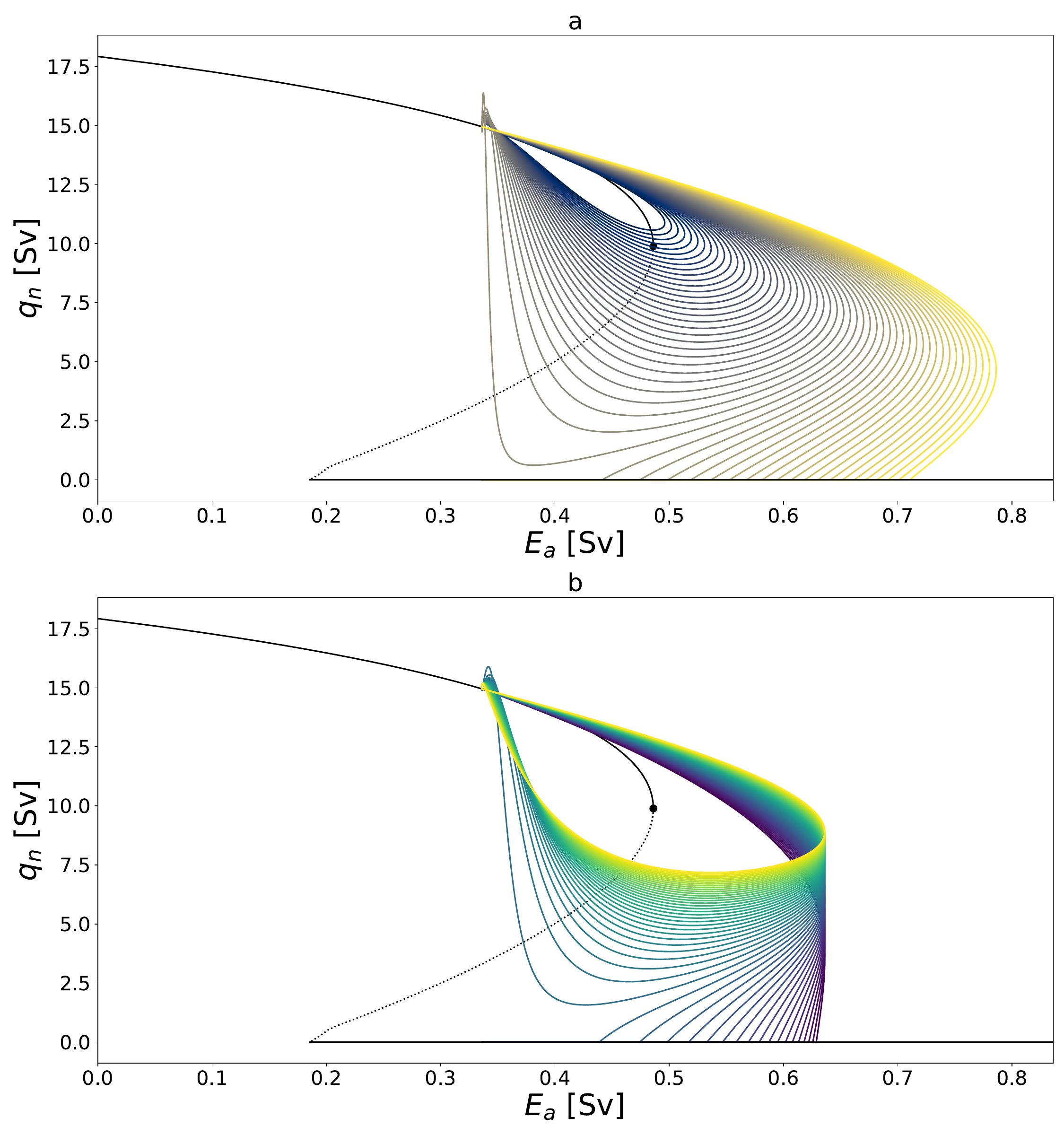}
    \caption{Bifurcation diagram of the AMOC strength $q_n$ with $E_a$ as bifurcation parameter for the parameter values in Table \ref{tab:par1} with stable states (solid, black) and saddle states (dashed, black), where the dot indicates the saddle-node bifurcation at which the ON-branch ends. (a) Several overshooting trajectories with constant freshwater flux $V_f = 3\e5$ km$^3$ (solid, blue to yellow for increasing $\varepsilon_a$), and (b) several overshooting trajectories with constant exceedance amplitude $\varepsilon_a = 0.15$ Sv (solid, purple to yellow for increasing rate $r$).}
    \label{fig:random}
\end{figure}

To illustrate the subtlety of a freshwater overshoot we have generated several overshooting trajectories. We use the model parameters from Table \ref{tab:par1} and fix $E_a^0 = 0.336$ Sv and $\Delta E_a = 0.15$ Sv. This way $E_a^0$ is roughly in the middle of the multiple equilibria regime. The size of the overshoot $\varepsilon_a$ is uniformly taken from the interval $[0.01\text{ Sv}, 0.30\text{ Sv}]$, and the associated forcing rate $r$ is taken such that the added freshwater volume $V_f$ is $3\e{5}$ km$^3$. In figure \ref{fig:random}a these trajectories are shown together with the bifurcation diagram in $(E_a, q_n)$ space. This illustrates that it is not necessarily the total added volume of freshwater that determines whether the AMOC collapses during an overshoot but also the rate at which this volume is deposited into the northern box. For low enough rates the trajectory spends enough time beyond the bifurcation point (i.e. the exceedance time) for the AMOC to collapse, while for sufficiently high enough rates the AMOC is too slow to respond and is still able to recover when the forcing returns to its original level, as has been discussed in \citet{ritchie2021overshooting}. For each exceedance amplitude $\varepsilon_a$ there is a critical rate $r_c$. A forcing rate falling short of $r_c$ will result in a collapsed AMOC, and hence a lower critical rate indicates a higher resilience of the AMOC to an overshoot.

\subsection{Physics of a freshwater overshoot}
\begin{figure*}
    \centering
    \includegraphics[width = 39pc]{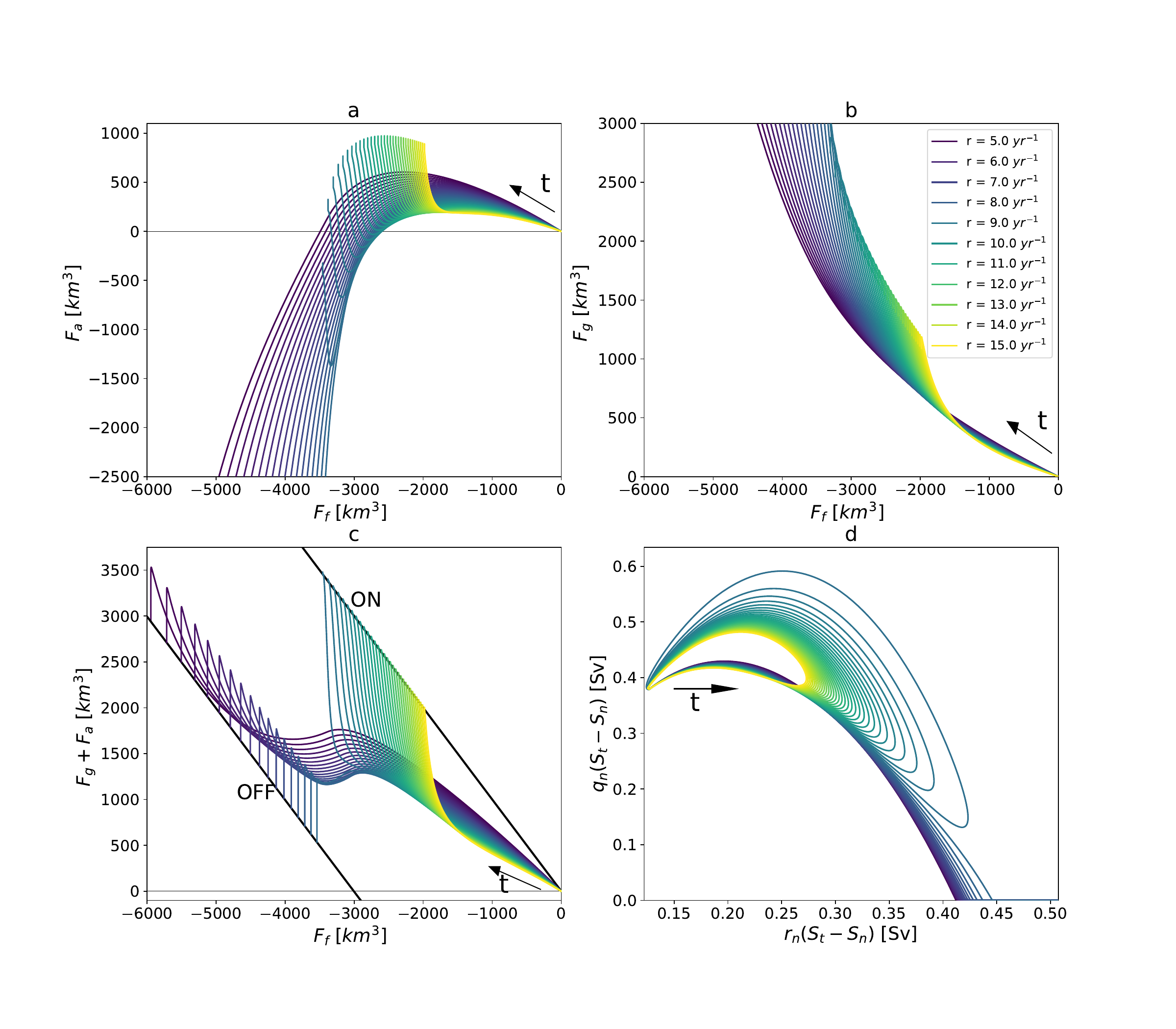}
    \caption{The cumulative fluxes (a) $F_a$ versus $F_f$, (b) $F_g$ versus $F_f$, (c) $F_g+F_a$ versus $F_f$ and (d) $q_n(S_t-S_n)$ versus $r_n(S_t-S_n)$ along several trajectories with $\varepsilon_a = 0.15$ Sv for various rates $r\in[5,15]$ yr$^{-1}$ (purple to yellow for increasing rate). The arrows indicate the direction of time $t$.}
    \label{fig:budget}
\end{figure*}

To determine which physical processes set this critical rate $r_c$ we generate under the same background parameters several trajectories for fixed exceedance amplitude $\varepsilon_a = 0.15$ Sv while the rate $r$ varies from $5$ yr$^{-1}$ to $15$ yr$^{-1}$, see figure \ref{fig:random}b. For this $\varepsilon_a$ the critical rate is $r_c \approx 8.53863$ yr$^{-1}$. During the overshoot mainly the salinity $S_n$ and temperature $T_n$ of the northern box vary, and as the salt-advection feedback is the mechanism that destabilizes the AMOC, we will focus on the salinity budget of the northern box during the overshoot. We define the cumulative added salinity by the freshwater flux $F_f$, the cumulative added salinity by the northern subtropical gyre $F_g$, and the cumulative added salinity by northward advective transport $F_a$ as
\begin{align*}
    F_f(t) &= \int_{-\infty}^t S_0\left(E_a^0 - E_a(t')\right)\,dt'\\
    F_g(t) &= \int_{-\infty}^t \big(r_n\left(S_t(t') - S_n(t') \right)-  r_n\left(S_t(0) - S_n(0) \right)\big)\,dt'\\
    F_a(t) &= \int_{-\infty}^t \big(q_n(t')\left(S_t(t') - S_n(t') \right) -\\
    &\quad\quad\quad\quad q_n(0)\left(S_t(0) - S_n(0) \right)\big)\,dt'.
\end{align*}
In other words, $F_f(t)$, $F_g(t)$ and $F_a(t)$ are the total volume of additionally imported salt fluxes up to time $t$ with respect to the starting ON state via either the surface freshwater flux, gyre transport or the advective AMOC transport, respectively. Figure \ref{fig:budget} depicts an overview of these fluxes for the generated trajectories.

From figure \ref{fig:budget}a it follows that initially the cumulative effect of an AMOC decline is a net salinification of the northern box. As the northern box is freshened by the forcing, the AMOC strength sharply declines since $q_n\propto S_n^2$. The remaining advective transport still imports the saline water from the tropics, but also exports the now freshened northern waters out off the northern box into the deep ocean, which causes this initial salinification. This net salinifying effect is less pronounced for higher rates of forcing as here the AMOC declines more abruptly. Along trajectories where the AMOC continues to decline the eventual lack of northward salt transport causes a net negative contribution to the salinity budget with respect to the stable ON state. A recovery is however still possible as there are several recovering trajectories with $F_a < 0$. Then following figure \ref{fig:budget}b where the cumulative gyre transport is shown, the dynamics are more straightforward: as $S_n$ declines while $S_t$ remains relatively constant the gyre transport increases. For higher rates of forcing, $S_n$ declines more sharply and so the gyre transport increases more abruptly as well. Hence for the same cumulative freshwater forcing $F_f$ the total volume of salt imported by the gyre is higher for higher forcing rates.

The combined effect of the gyre and advective terms are shown in figure \ref{fig:budget}c together with the line $F_f = F_g + F_a$ representing the ON state as all the added freshwater flux $F_f$ has been removed by the gyre and advective transport. In a similar way the line $F_f = F_g + F_a - V_n\left(S_n^{\text{ON}}-S_n^{\text{OFF}}\right)$ represents the OFF state where all added freshwater $F_f$ has been removed by the gyre and advective term up to an amount $V_n\left(S_n^{\text{ON}}-S_n^{\text{OFF}}\right)$, which is the difference in freshwater content of the northern box between the ON and OFF state. Initially for the trajectories forced under higher rates less of the added freshwater has been removed. However, as for those the salinity $S_n$ drops quicker, the gyre transport picks up sooner which saves the AMOC. For those with lower forcing rates the decline of $S_n$ is slower and the lingering advective transport even has initially an additional salinifying effect, but as the gyre transport does not pick up quickly enough now, the northern salinity continues to drop and eventual the AMOC collapses. Lastly, in figure \ref{fig:budget}d the advective transport is plotted against the gyre transport. This indeed shows that for higher forcing rates the gyre transport is greater for the same amount of advective transport as the trajectories start to recover. This distinguishes the recovering trajectories from the collapsed ones. Note in this figure that the recovering trajectories make an anti-clockwise loop. This is due to an increase in $S_t$ as the northward transport out of this box was temporarily halted. All in all, the gyre transport is the main factor determining an AMOC collapse during an overshoot. When the forcing rate is higher, the salinity of the northern box drops more abruptly, which results in a higher gyre transport with these trajectories for the same cumulative added freshwater volume $F_f$. 

\begin{figure*}
    \centering
    \includegraphics[width = 39pc]{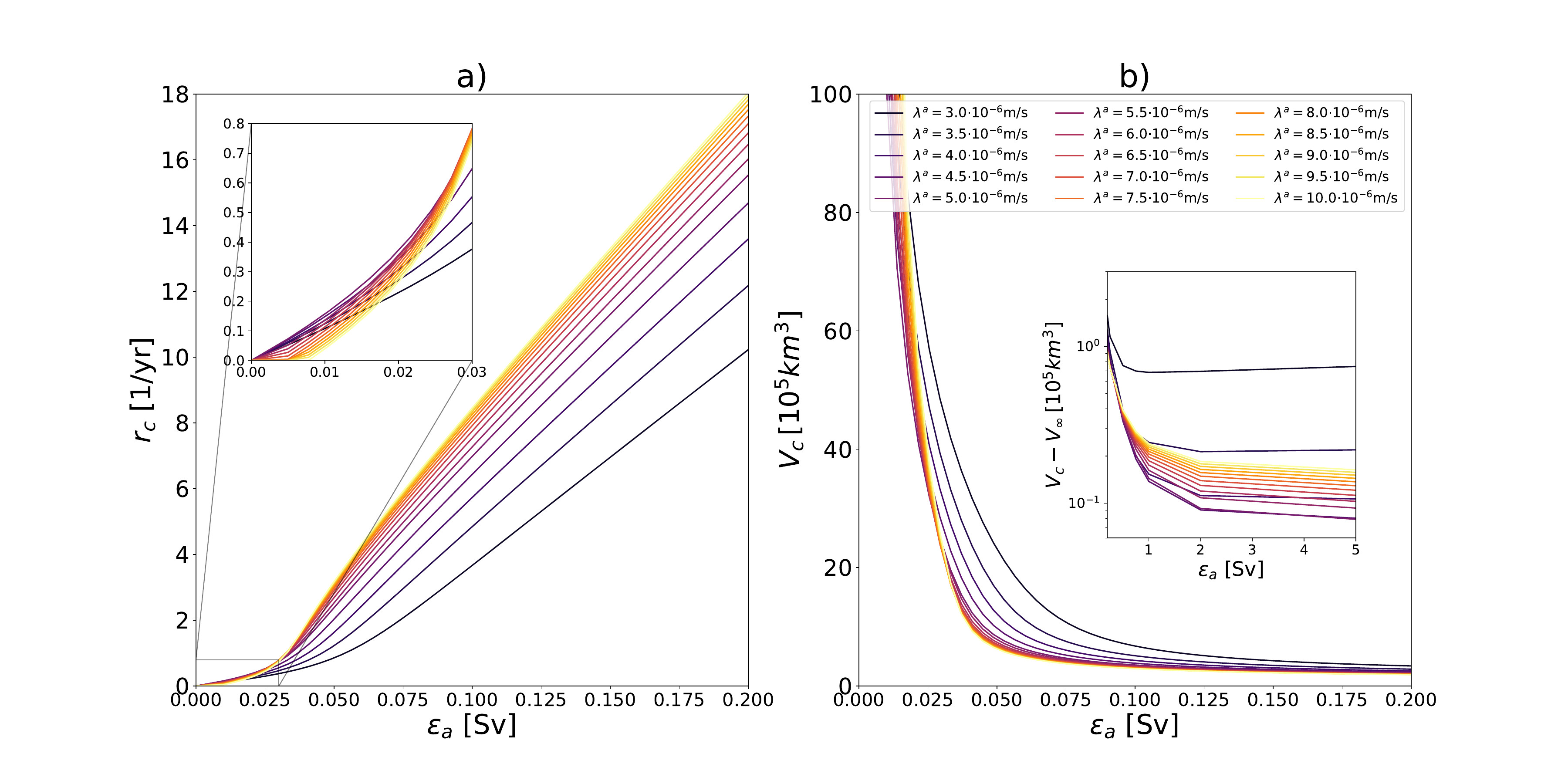}
    \caption{The critical rate $r_c$ (a) and associated added freshwater volume $V_c$ (b) for exceedance amplitudes $\varepsilon_a\in[0 \text{ Sv},0.2\text{ Sv}]$ for various atmosphere-ocean coupling parameters $\lambda^a$ (black to yellow for increasing value) with the inset in (b) indicating the difference $V_c - V_\infty$ for $\varepsilon_a\in[0.2\text{ Sv},5\text{ Sv}]$.}
    \label{fig:rc}
\end{figure*}

\subsection{Effect of the exceedance amplitude}
The resilience of the AMOC to an overshoot is analyzed for various exceedance amplitudes $\varepsilon_a\in[0.005 \text{ Sv}, 0.20\text{ Sv}]$ and atmosphere-ocean heat exchange parameters $\lambda^a\in\left\{3.0\e{-6}\text{ m/s},\, 3.5\e{-6}\text{ m/s},\dots, \,10.0\e{-6}\text{ m/s}\right\}$. Note that as the latter varies the atmospheric temperatures of the boxes also vary (see Table \ref{tab:par2}) in order to keep the state variables of the unforced ON state at $E_a = 0$ Sv unchanged. However, the bifurcation structure does change as with increasing $\lambda^a$ the multiple equilibria regime occurs for lower freshwater forcing values. This is caused by the diminished stabilizing thermal feedback since the temperatures of the surface boxes follow the atmospheric temperatures more closely. We keep the distance to the bifurcation point $\Delta E_a = 0.15$ Sv fixed. As an indication of the AMOC's resilience to overshooting we use the critical rate $r_c$ and the associated critical total freshwater input $V_c$, following \eqref{eq:vf}, which are shown in figure \ref{fig:rc}. Here, $V_c$ is the minimal total freshwater input via the forcing protocol \eqref{eq:forcing0} that still leads to AMOC tipping. For each $\lambda^a$ we have added the origin as additional point as $r_c \downarrow 0$ for $\varepsilon_a\downarrow0$. 

We see indeed that the critical rate increases --and hence the resilience decreases-- as the exceedance amplitude of the freshwater forcing $\varepsilon_a$ increases: a larger freshwater perturbation needs to end earlier in order to save the AMOC. Moreover, note that for each coupling parameter $\lambda^a$ we can distinguish two regimes. The first is characterized by small forcing rates ($r_c \lessapprox 2$ yr$^{-1}$) where $r_c$ grows super-linearly with the overshoot $\varepsilon_a$, as can be seen in the inset of figure \ref{fig:rc}a. The second regime sees large forcing rates ($r_c \gtrapprox 2$ yr$^{-1}$) that increase linearly with $\varepsilon_a$. In the former, the forcing timescale ($1/r_c$) is large relative to the system's response time, and hence other boxes besides the directly affected northern and southern box adapt during the overshoot as well. In the latter regime, however, the forcing becomes pulse-like as its time scale approaches zero $1/r_c\to 0$, and the AMOC is shut down almost instantaneously by a sharp drop in the northern salinity. 

\begin{figure*}
    \centering
    \includegraphics[width = 39pc]{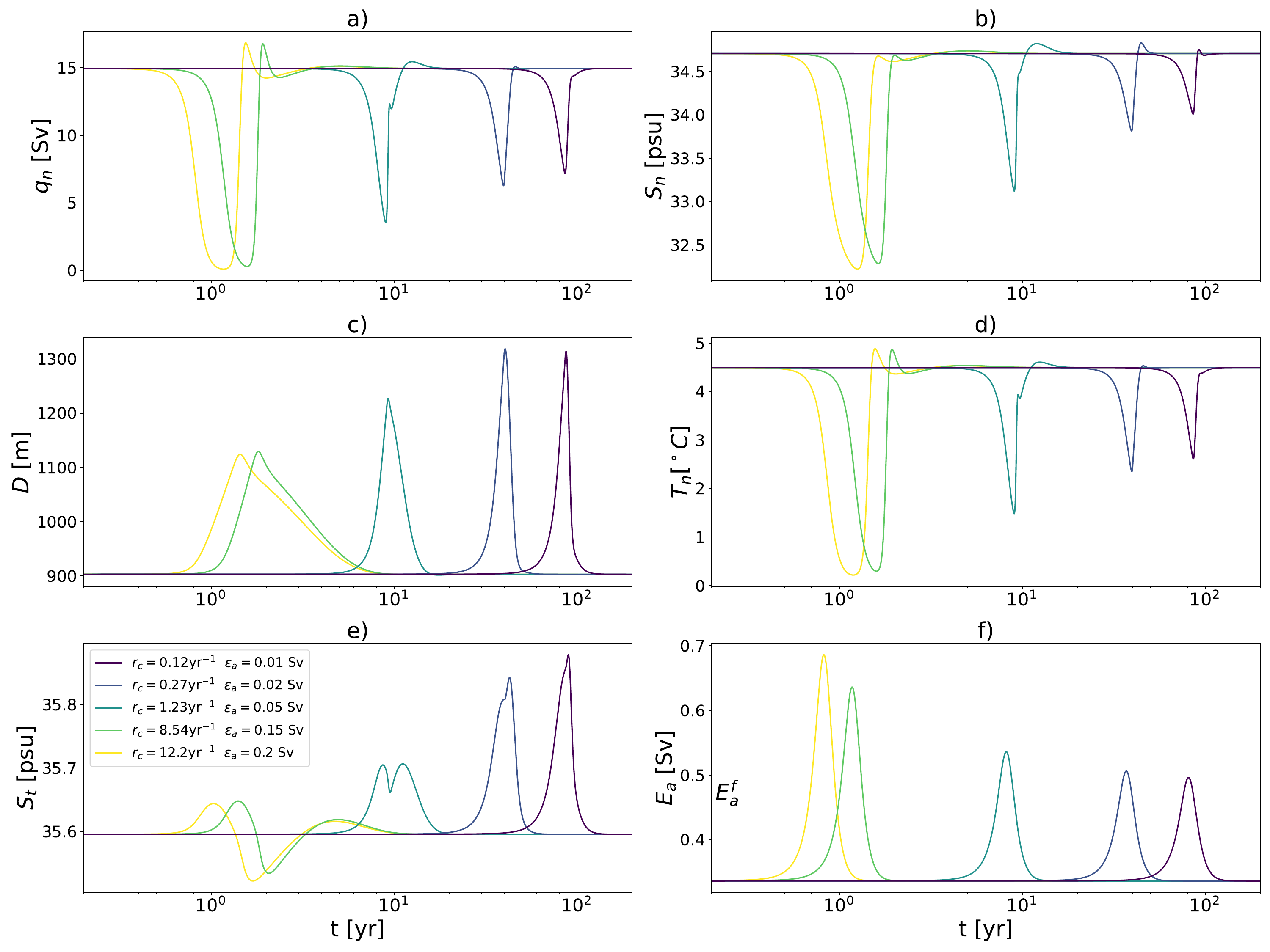}
    \caption{The AMOC strength $q_n$ (a), pycnocline depth $D$ (b), salinity $S_n$ (c) and temperature $T_n$ (d) of the northern box, salinity $S_{t}$ of the $t$-box (e) and salinity $S_{ts}$ of the $ts$-box (f) of overshooting trajectories where $r\lessapprox r_c$ for various exceedance amplitudes $\varepsilon_a\in[0.01\text{ Sv},0.18\text{ Sv}]$ (increasing with purple to yellow) with $\lambda^a = 3.5\e{-6}$ m/s. Note the logarithmic scaling of the time axes.}
    \label{fig:laconstant}
\end{figure*}

We first discuss this linear regime. Here the pulse-like forcing injects freshwater into the northern box and the AMOC strength $q_n$ drops almost immediately to zero, see figure \ref{fig:laconstant}a. In this figure we fix $\lambda^a = 3.5\e{-6}$ m/s and consider among others two cases with a critical rate $r_c > 2$ yr$^{-1}$ with $r_c \approx 8.54$ yr$^{-1}$ and $r_c \approx 12.2$ yr$^{-1}$ corresponding to an overshoot of $\varepsilon_a = 0.15$ Sv and $\varepsilon_a = 0.2$ Sv respectively. These only exceed the bifurcation level for $0.31$ yr and $0.25$ yr, respectively. The actual applied forcing rates are marginally larger than $r_c$ so the AMOC only just recovers. Note furthermore that only $T_n$ and the pycnocline depth $D$ react substantially during the AMOC decline, while $S_t$ shows minimal change, mainly due to the change in volume $V_t$. Variables not directly affected by the freshwater pulse remain relatively unchanged as the forcing is applied. As the rates and exceedance amplitudes increase, these forcings become (approximately) instantaneous and hence the gyre and advective transports have less time to remove the perturbation. Eventually, for large enough rates, just a fixed volume is needed that drops the northern salinity to a level where $q_n \leq 0$ Sv. This is consistent with the linear relationship $r_c\propto \varepsilon_a$ which implies that $V_c$ will asymptotically approach a constant value $V_\infty$ as $\varepsilon_a\to\infty$. That the AMOC can still recover despite being almost zero is due to the sufficiently high enough salinity $S_t$ in the tropical box. As the AMOC only slightly increases, it transports this salt northward, which aides its recovery via the salt-advection feedback.

Given these dynamics, we can now formulate an expression for this theoretical limit $V_\infty$. We inject a volume $V_\infty$ of freshwater in the northern box and assume that the northern salinity changes instantaneously to
\begin{equation*}
    S_n^{\text{ON}} \to S_n^{\text{ON}} - S_0\frac{V_\infty}{V_n}.
\end{equation*}
This assumption only holds in the limit $r_c\to\infty$, as for any finite forcing rate there is time for the gyre and advective transports to remove the injection partly. Note that then in this limit the previously described physical processes of an AMOC overshoot are no longer valid. Under this instantaneous change the AMOC strength drops to
\begin{equation}\label{eq:Vinf}
    q_n \to \eta D^2 \left(\alpha\left(T_{ts}^{\text{ON}}-T_{n}^{\text{OFF}}\right) - \beta \left(S_{ts}^{\text{ON}}-\left(S_n^{\text{ON}} - S_0\frac{V_\infty}{V_n}\right)\right)\right)
\end{equation}
where it is taken that the northern temperature drops with the AMOC as well, as can been seen in figure \ref{fig:laconstant}d. As $V_\infty$ is a volume just large enough to drop the AMOC strength to zero, solving for $V_\infty$ by equating \eqref{eq:Vinf} to zero yields:
\begin{equation}
    V_\infty = \frac{V_n}{S_0}\left(S_n^{\text{ON}}-S_{ts}^{\text{ON}}+\frac{\alpha}{\beta}\left(T_{ts}^{\text{ON}}-T_{n}^{\text{OFF}}\right)\right).
\end{equation}
This theoretical limit is compared to the computed results, see the inset of figure \ref{fig:rc}b. It shows that indeed the critical volumes do approach their respective limit $V_\infty$, but that there is still a persistent difference as $T_n$ does not immediately cool down to $T_{n}^{\text{OFF}}$ especially for the lower values of $\lambda^a$. The relative error levels off between $5.7\%$ and $29.0\%$ with median at $10.8\%$ for $\varepsilon_a = 5$ Sv. Hence this theoretical limit is only a reasonable approximation for values outside of the relevant domain. 

Next, we discuss the second regime. Here the critical forcing rates are low and grow super-linearly with the exceedance amplitude. In figure \ref{fig:laconstant} three cases are shown with respective overshoots $\varepsilon_a$ of $0.01$ Sv, $0.02$ Sv and $0.05$ Sv with relatively low critical rates, and so the forcing exceeds the bifurcation for $58$ yr, $21$ yr and $3.4$ yr respectively. In these cases the AMOC strength $q_n$ and northern salinity and temperature do not drop as low, although the total amount of freshwater added is much larger. Here the gyre has more time to remove the freshwater perturbation. Moreover, the salinity added to the southern box now has time to move northward resulting in a higher peak in $S_t$. Note that $S_t$ in case of $\varepsilon_a = 0.05$ has a double peak: the minimum is caused by the pycnocline peaking resulting in a larger volume $V_t$. The pycnoclines now also attain larger maxima as $q_n$ is halted for a longer duration. Now, in these cases the AMOC barely recovers, but its minimum is still at respectively $7.2$ Sv, $5.7$ Sv and $3.6$ Sv, instead of the previous cases that were below $0.4$ Sv. So in these low forcing rate cases the AMOC is still substantial during these critical trajectories compared to those in the high forcing rate cases. We can intuitively explain why the AMOC does not attain lower values here, but reaches almost $0$ Sv for the higher values of $\varepsilon_a$: these low AMOC strengths are only attained  for less than a year, and so the northern box only \say{misses out} on northward salt transport for a short time. In contrast, for the former cases, where $r_c\to0$, the trajectories are close to equilibrium, and low AMOC strengths have to be sustained for a longer time interval. Note that in the true limit $r_c\to 0$ the system will reach the bifurcation point, and the lowest AMOC strength that can be attained is $q_n^{\text{ON}}\big|_{E_a^b}$, which is $9.9$ Sv for this scenario.

It remains to discuss why here $r_c$ increases initially super-linearly with $\varepsilon_a$. This trend can be explicitly computed using the approximation formula found by \citet{ritchie2019inverse}, which holds for small $r_c$ and $\varepsilon_a$. However, this is only valid when the stable branch ends in a saddle-node bifurcation, which is solely the case for $\lambda^a\in\{3.0\e{-6}\text{ m/s},\, 3.5\e{-6}\text{ m/s},\, 4.5\e{-6}\text{ m/s},\,5.0\e{-6}\text{ m/s}\}$. Hence we omit an explicit expression, and only argue the trend heuristically. We can safely assume $\der{}{\varepsilon_a}\left(\frac{\pi\varepsilon_a}{r_c}\right)<0$, otherwise we would be able to add more freshwater at a higher flux past the bifurcation without an AMOC collapse. This implies that $r_c$ should at least grow linearly with $\varepsilon_a$. Now, for small forcing rates an incremental increase in the exceedance amplitude needs not only be met with a proportional (linear) increase in forcing rate such that the exceedance volume does not grow, but also by an additional increase in the forcing rate since some stabilizing mechanisms are less effective for these higher rates. An example of this can be seen in figure \ref{fig:laconstant}: as the critical rates increase, the peak in the tropical salinity $S_t$ declines, as other state variables now also can adapt to a weakened AMOC during these lengthy overshoots. An increased tropical salinity aides the AMOC's stability since it enhances the northward salt transport. Another example is the temperature of the deep box ($T_d$, not shown): for low forcing rates the deep ocean can release more heat during the overshoot, with $T_d$ dropping to $3.1^\circ$C for the low rate $r_c\approx 0.12$yr$^{-1}$ while not getting below $4.4^\circ$C for the high forcing rate $r_c\approx 12.2$yr$^{-1}$. This heat is released to the Southern Ocean, decreasing its density, which in turn aides the AMOC strength. Note moreover that the two cases in the linear regime are dynamically not as distinct: in all state variables the trajectory for $\varepsilon_a = 0.2$ Sv seems to be a sped up version of the case $\varepsilon_a = 0.15$ Sv. So in the linear regime there does not seem to be an additional loss of a stabilizing mechanism. As the forcing becomes pulse-like, the dynamical response stays similar. To conclude, for small exceedance amplitudes the critical forcing rates initially grow super-linearly. In this small $\varepsilon_a$ regime the dynamical response of the system changes with the amplitude, as for example the smaller peaks in $S_t$ and $T_d$ illustrate. This means that on top of a linear increase in the critical forcing rate consistent with a non-decrease in added freshwater volume, there is additional loss of resilience, and hence super-linear growth of the critical rates. In the high $\varepsilon_a$ regime on the other hand there is no qualitative change in the system's dynamics with increasing amplitude, as for these high critical rates the forcing is practically instantaneous. Hence only the added freshwater volume matters and so the rates only increase linearly.

\subsection{Effect of the atmosphere-ocean coupling}
Regarding the atmosphere-ocean coupling, we discuss the AMOC's resilience to an overshoot in relation to the coefficient $\lambda^a$. In the linear regime the critical forcing rates $r_c$ are larger for larger $\lambda^a$. This follows as a larger atmosphere-ocean coupling coefficient causes a weaker thermal feedback mechanism and so decreases AMOC resilience. The thermal feedback entails that a weakening of the AMOC comes with a diminished northward heat transport, resulting in an increased density in the north, which in turn strengthens the AMOC. However, a more efficient heating by the atmosphere can partly compensate for this heat transport loss. A stronger coupling also means that $V_\infty$ is smaller, i.e. less freshwater is needed to instantaneously cause a collapse as $T_n^{\text{OFF}}$ is larger for increased coupling. Since for $r_c\propto\varepsilon_a$ we can derive that $\derp{r_c}{\varepsilon_a}\to \frac{\pi}{V_\infty}$ as $\varepsilon_a\to\infty$, it follows that the linear growth rates with $\varepsilon_a$ should indeed also increase with larger $\lambda^a$. 

More interesting is the AMOC's resilience to small overshoots, see the inset in figure \ref{fig:rc}a. Here we can see that a larger coupling parameter actually increases the AMOC's resilience as $\varepsilon_a\downarrow0$. The growth rate with $\varepsilon_a$ is however also greater. In this regime the overshooting trajectory can be viewed as a quasi-equilibrium trajectory instead of a transient one. Now, the ON-equilibria are more resilient to a freshwater perturbation for larger $\lambda^a$. This can be deduced from the fact that $-\der{q_n^{\text{ON}}(E_a)}{E_a}\big|_{E_a^b}$ is larger for smaller coupling coefficient. This also implies that the $(E_a,q_n)$ bifurcation diagram is steeper when approaching the bifurcation. If we compute $\left(q_n^{\text{ON}}(E_a^b-\delta E_a)- q_n^{\text{ON}}(E_a^b)\right)/\delta E_a$ for $\delta E_a = 1$ mSv, we obtain $1504.1$ for $\lambda^a = 3.0\e{-6}$ m/s which decreases to $79.2$ for $\lambda^a = 10.0\e{-6}$. This can also be explained physically: for lower coupling coefficients the AMOC is less thermally driven and more haline driven, since the atmosphere cannot enforce a large temperature difference as easily. This means that the AMOC is more sensitive to freshwater perturbations. Now, as $\varepsilon_a$ and so $r_c$ increase, the AMOC strength and $T_n$ will drop further and quicker, see also figure \ref{fig:laconstant}a \& d. Hence, the thermal feedback mechanism starts to play a more significant role during an overshoot, which yields that $r_c$ will increase with a higher $\lambda^a$, as here the drop in $T_n$ will not be as severe.  

Note that the differences in critical rates may seem small, but the resulting differences in exceedance time and volume can be huge. For example at $\varepsilon_a = 0.01$ Sv we have $r_c = 0.11$ yr$^{-1}$ for $\lambda^a = 3.0\e{-6}$ m/s but $r_c = 0.04$ yr$^{-1}$ for $\lambda^a = 10.0\e{-6}$ m/s. This implies that under the higher coupling coefficient one could exceed the bifurcation threshold for an additional 110 years and add an additional $5\e{5}$ km$^3$ of freshwater to the northern box without an AMOC collapse.


\begin{figure*}
    \centering
    \includegraphics[width = 39pc]{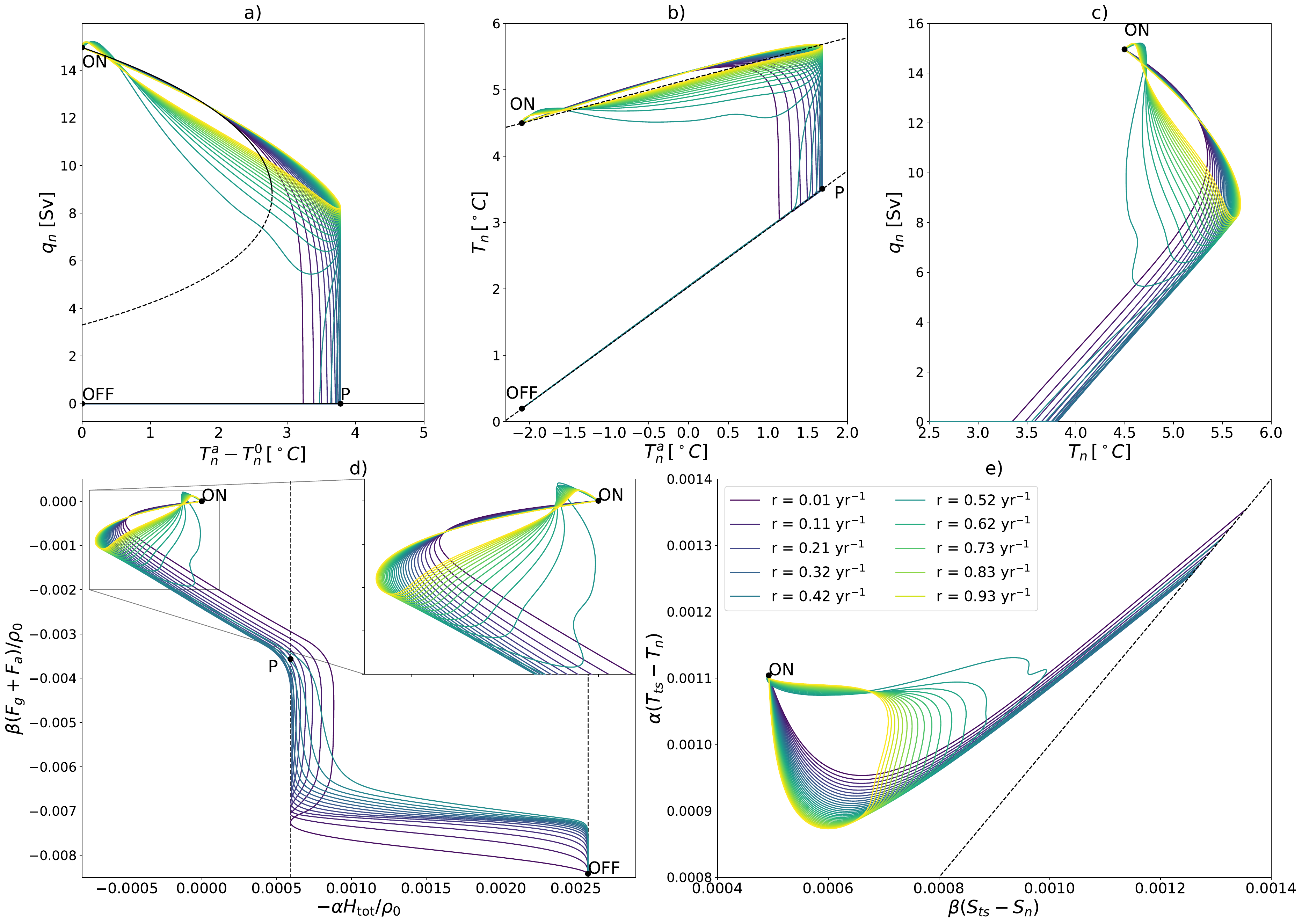}
    \caption{The AMOC strength $q_n$ versus the rise in atmospheric temperature $T_n^a - T_n^0$ together with the underlying bifurcation diagram (a), the northern box's temperature $T_n$ versus the actual atmospheric temperature $T_n^a$ (b), the AMOC strength $q_n$ versus the box's temperature $T_n$ (c), the northern box's non-dimensionalized density increase due to salinity $\beta(F_g+F_a)/\rho_0$ versus this increase due to heat $-\alpha H_{\text{tot}}/\rho_0$ (d), and the non-dimensional density difference due to temperature versus salinity (e) all for various overshooting trajectories with fixed amplitude $\theta_a = 1\,^\circ$C and varying rate $r\in[0.01\text{ yr}^{-1}, 1\text{ yr}^{-1}]$ (increasing with purple to yellow). When needed, the dots indicate the original starting ON-state, the OFF-state under the original starting temperature $T_n^0$, and dot $P$ the OFF-state under the maximum atmospheric temperature $T_n^0 + \Delta T_n^a + \theta_a$. The dashed lines in (c) indicate the approximate relation of $T_n$ versus $T_n^a$ that should hold given all other variables are in either the starting ON or ending OFF-state. Lastly, in (d) the leftmost dashed vertical line indicates the net density increase via heat transports needed to reach state P, and the rightmost dashed vertical line indicates this amount needed to reach the final OFF-state, while in (e) the dashed line indicates where the AMOC strength would be zero.}
    \label{fig:tempbif}
\end{figure*}

\section{Overshoot in Atmospheric Temperature}\label{sec:temp}
Now we treat the northern atmospheric temperature $T_n^a$ as our overshooting parameter. We alter $T_n^a$ with a similar forcing profile as for the freshwater forcing:
\begin{equation}\label{eq:forcing}
    T_n^a(t) = T_n^0+ \frac{\Delta T_n^a + \theta_a}{\cosh{(rt)}},
\end{equation}
where $T_n^0$ is the initial temperature of the equilibrium ON-state with a distance $\Delta T_n^a$ to the bifurcation threshold. Hence $\theta_a$ is the exceedance amplitude of the temperature overshoot. We fix $\lambda^a = 3.5\e{-6}$ m/s, since we do not expect the dynamics in a temperature overshoot to change qualitatively by varying the coupling parameter. The atmospheric overshoot would just be enforced more quickly onto the ocean. The initial temperature $T_n^0$ is then set to $-2.0965\, ^\circ$C (see Table \ref{tab:par2}), and the freshwater forcing is fixed at $E_a = 0.336$ Sv, so a comparison with the previous results can be directly made. This results in a saddle-node bifurcation at $T_n^a = 0.683 \, ^\circ$C, and so $\Delta T_n^a = 2.779\, ^\circ$C. 

\subsection{Physics of a temperature overshoot}
In the same manner as before, we first fix the exceedance amplitude $\theta_a = 1\,^\circ$C and apply a rate $r\in[0.01\text{ yr}^{-1}, 1\text{ yr}^{-1}]$. Again an AMOC collapse is determined by the rate of forcing, where a rate smaller than the critical rate $r_c \approx 0.52$ yr$^{-1}$ results in a collapsed AMOC. Under these forcings the increase in temperature of the northern box $T_n$ is persistent enough such that an active AMOC can no longer be sustained. The results are shown in figure \ref{fig:tempbif}.

In figure \ref{fig:tempbif}a the bifurcation diagram in $(T_n^a - T_n^0, q_n)$-space is shown together with the overshooting trajectories. Several trajectories forced with a low rate already exhibit an AMOC collapse before the maximum atmospheric temperature is reached, while several of the recovering trajectories show still a significant decrease in the AMOC strength. We have indicated the starting ON state and its corresponding OFF state, together with the OFF state associated with the maximum atmospheric temperature $T_n^0 + \Delta T_n^a + \theta_a$ indicated with $P$. The collapsed trajectories with a relatively low forcing rate pass close to this state. Note that the only direct physical impact of the atmospheric temperature changes onto the AMOC is the resulting temperature $T_n$ of the northern box, and so figure \ref{fig:tempbif}b exhibits the atmospheric temperature and the corresponding oceanic temperature of the northern box. The two dashed lines indicate a linear approximation for $T_n$ as a function of $T_n^a$ under the assumption that the heat budget of the northern box is in equilibrium, where the top line assumes an active AMOC while the bottom is for a collapsed circulation. These approximations are derived by assuming
\begin{align*}
    V_n\der{T_n}{t} = \left(q_n+r_n\right)\left(T_t - T_n\right) - \lambda^a A_n\left(T_n-T_n^a\right) \approx 0 
\end{align*}
in case of an active AMOC. Then differentiating with respect to $T_n^a$ while we suppose that the other state variables do not depend on the atmospheric temperature, yields
\begin{align*}
    &-2\alpha\eta D^2T_n\der{T_n}{T_n^a}\\
    &\,\,+\left(r_n + \lambda^a A_n + \eta D^2(\alpha T_{ts}-\beta(S_{ts}-S_n))\right)\der{T_n}{T_n^a} = \lambda^a A_n
\end{align*}
and rearranging produces our linear approximation
\begin{align*}
    &\der{T_n}{T_n^a}\Bigg|_{\text{ON}} = \left(\lambda^a A_n\right)/\Big(-2\alpha\eta (D^{\text{ON}})^2T_n^{\text{ON}} + r_n  \\
    &\,\,+ \eta (D^{\text{ON}})^2(\alpha T_{ts}^{\text{ON}}-\beta(S_{ts}^{\text{ON}}-S_n^{\text{ON}}))+ \lambda^a A_n \Big).
\end{align*}
This shows that -- as expected -- $T_n$ follows the atmospheric temperature more closely for increased ocean-atmosphere coupling, while this tracking is also influenced by terms in the denominator. The first term represents the thermal feedback effect, which aides this tracking as an increase in $T_n$ would decrease the northward advective transport of heat. The second and third term represent the gyre transport and the part of the AMOC transport that is independent of the northern temperature, respectively, which both inhibit the effect of $T_n^a$ as they transport heat northward. A similar approximation can be done for a collapsed circulation, yielding
\begin{align*}
    \der{T_n}{T_n^a}\Bigg|_{\text{OFF}} = \left(\lambda^a A_n\right)/\left( r_n + \lambda^a A_n \right),
\end{align*}
which is just the ratio of the rate of the atmospheric heat flux to the rate of all heat fluxes. Now, in figure \ref{fig:tempbif}b the overshooting trajectories track these approximations reasonably well for an active AMOC, and almost exactly for a collapsed circulation. This implies that during the overshoot the northern box's heat budget is almost always in approximate equilibrium, except during the AMOC collapse or for a weakened AMOC. During such a weakening or collapse the increase in atmospheric temperature is opposed by the decrease in northward heat transport, yielding a net cooling of the northern box, as can be seen in figure \ref{fig:tempbif}c. The largest rise in $T_n$ is reached for the largest forcing rate where there is the least AMOC weakening.

Now, to investigate the causes of this weakening we turn to figure \ref{fig:tempbif}d, where we plot the non-dimensional increase in density of the northern box due to the salinity variations by the changing advective and gyre transports on the ordinate, which is given by $\beta\left(F_a + F_g\right)/\rho_0$. We also define similar fluxes as previously but now for heat transports:
\begin{align*}
    H_f(t) &= \lambda^a A_n\int_{-\infty}^t \left(\big(T_n(t') - T_n^a(t')\big) - \big(T_n(0) - T_n^0\big)\right)\,dt'\\
    H_g(t) &= \int_{-\infty}^t \big(r_n\left(T_t(t') - T_n(t') \right)  -r_n\left(T_t(0) - T_n(0) \right)\big)\,dt'\\
    H_a(t) &= \int_{-\infty}^t \big(q_n(t')\left(T_t(t') - T_n(t') \right) -\\
    &\quad\quad\quad\quad q_n(0)\left(T_t(0) - T_n(0) \right)\big)\,dt'
\end{align*}
with total cumulative additional heat transport into box $n$ as $H_{\text{tot}} = H_f + H_a + H_g$. Therefore the net non-dimensional density input due to temperature by the advective and gyre transports, and the atmospheric forcing follows as $-\alpha H_{\text{tot}}/\rho_0$. The figure shows that initially the density changes are dominated by the temperature effects as the northern box heats up. This in turn weakens the AMOC, which causes a loss of salinity transport and heat transport, where the former causes the actual collapse via the salt-advection feedback. It can be deduced from the figure that the density changes due to salinity outweigh those due to temperature. The difference between a recovering and collapsing trajectory is that the latter have a larger density loss in salinity for the same density change in temperature. As they collapse further, the decrease in $T_n$ overcompensates for the initial density loss due to the heating. As they collapse, they pass $P$, where the salinity fluxes adjust quickly to the new collapsed circulation. From state $P$ onward to the final OFF state the density changes are then dominated by temperature changes. For the recovering trajectories we also see that the cooling induced by an AMOC weakening can completely compensate for the initial temperature forcing, and that the recovery is also mainly driven by the salt-advection feedback. Hence, in both cases the initial temperature forcing is only a small perturbation in terms of density variations.

The difference between a collapse and a recovery is that in the former the AMOC is weakened long enough such that the loss in salinity is large enough to instigate a collapse via the salt-advection feedback. This is shown more clearly in figure $\ref{fig:tempbif}$e, where the non-dimensional density differences driving the AMOC due to salinity ($\beta \left(S_{ts}-S_{n}\right)$) versus temperature ($\alpha \left(T_{ts}-T_{n}\right)$) are depicted. Safe overshoots have for the same density deficit caused by temperature a smaller density deficit in salinity than unsafe overshoots. A high rate of temperature forcing can have the same impact as a low rate on the temperature-induced density difference, but critical in the end is the difference in salinity, which --via the salt-advection feedback-- causes the collapse. Whether this salt-advection feedback takes over, follows via the same mechanism as for the freshwater overshoot: an initially steeper drop in $S_n$ will activate the gyre and add additional salt from the south. 

\begin{figure*}
    \centering
    \includegraphics[width = 39pc]{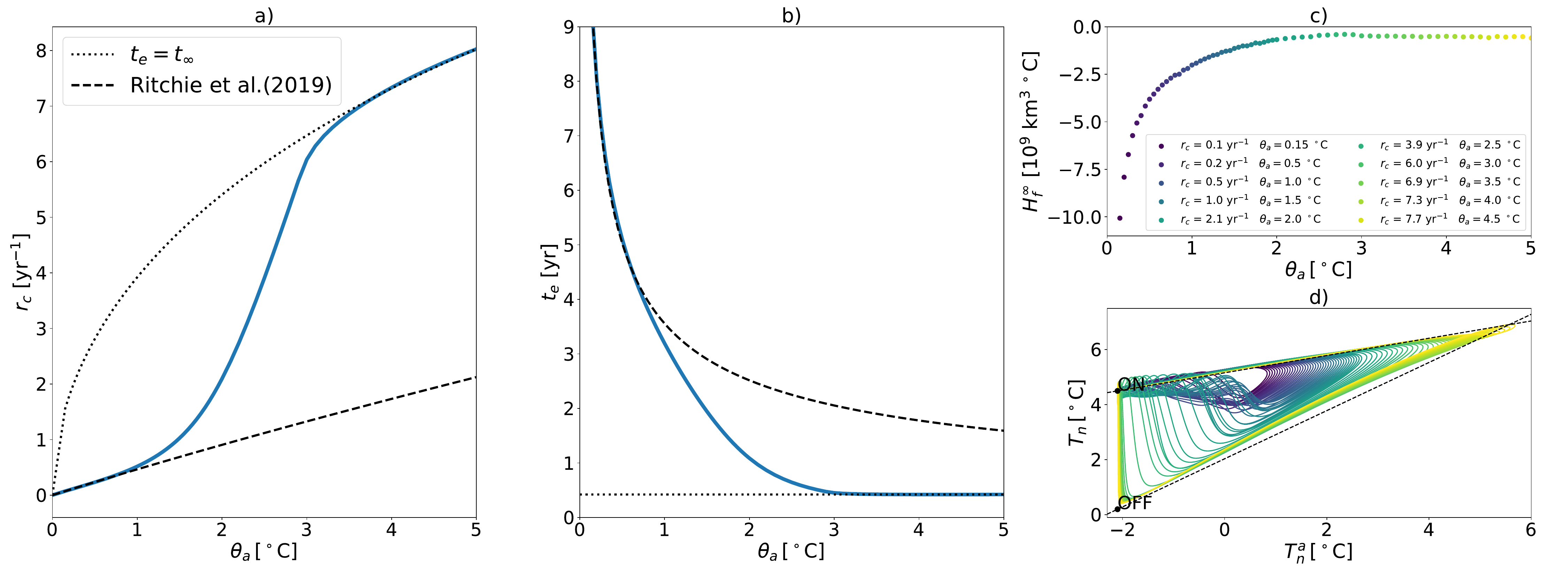}
    \caption{The critical rates $r_c$ (a) and exceedance time $t_e$ past the bifurcation point (b) for varying exceedance amplitude $\theta_a$ with the fits following \citet{ritchie2019inverse} (dashed) and assuming fixed exceedance time $t_\infty$ (dotted). For various exceedance amplitudes overshooting trajectories with forcing rate $r\lessapprox r_c$ (purple to yellow for increasing amplitude) the total atmospheric heating $H_f^\infty$ (c) and $T_n$ versus $T_n^a$ (d) where the dashed lines represent the same approximations as in figure \ref{fig:tempbif}b.}
    \label{fig:temprc}
\end{figure*}

\subsection{Effect of the exceedance amplitude}
Next, we vary the exceedance amplitude $\theta_a\in[0.15\,^\circ\text{C}, 5\,^\circ\text{C}]$ which corresponds to a maximum total rise in atmospheric temperature of $8^\circ\text{C}$. The associated critical rates $r_c$ and exceedance times past the bifurcation point $t_e$ are shown in figure \ref{fig:temprc}a \& b. Similar to the freshwater overshoot we can distinguish two regimes: a regime where $r_c\propto \theta_a$ for small amplitudes $\theta_a\lessapprox 1\,^\circ$C, and one where $r_c\propto \log(\theta_a)$ for $\theta_a\gtrapprox 3.5\,^\circ$C with a transitioning regime in-between. The corresponding exceedance times show a similar trend as in figure \ref{fig:rc}b: rapidly descending and then asymptotically approaching a limit value $t_\infty \approx 0.4$ yr. Figures \ref{fig:temprc}a \& b additionally show two approximations: one for the limit of small exceedance $\theta_a\downarrow0$ following \citet{ritchie2019inverse} and one assuming a constant exceedance time $t_\infty$, both showing a good fit for both limit regimes. 

The high $\theta_a$ regime can be explained in a similar way as the large-amplitude freshwater overshoot: for a large enough rate the dynamical response of the system is too slow and only the net density reduction of the northern box matters. Indeed we see that the total atmospheric heating, $\lim_{t\to\infty}H_f(t) := H_f^\infty$, approaches a limiting value (figure \ref{fig:temprc}c). This limit process corresponds to the exceedance time $t_e$ approaching $t_\infty$, since for larger temperature overshoots $t_e$ only reduces slightly in order to still have the same heating flux $H_f^\infty$. Note we have limited ourselves to amplitudes $\theta_a\leq 5\,^\circ$C, as for higher values the linear approximations in figure \ref{fig:temprc} cross. This means that for these high values of $T_n^a$ the quasi-equilibrium value of $T_n$ for a ON-state is lower than $T_n$ of a the OFF-state, i.e. for these large northern atmospheric temperatures the AMOC would actually have a cooling effect on the northern ocean temperature. As this is nonsensical, we have limited our amplitudes. Moreover, figure \ref{fig:temprc}d shows that for larger rate values the recovering trajectories also see a large drop in $T_n$ and pass along the equilibrium OFF-line. Similar to the freshwater case we have comparable dynamics for large amplitudes.  The forcings close to the critical rate cause the AMOC to almost immediately drop to zero, and with it $D$, $S_n$ and $T_n$ drop, while the other state variables are relatively unperturbed.

For the low $\theta_a$ regime we also observe the same dynamical response as for the small freshwater overshoots. Under lower forcing ratings the system has more time to adjust to the weakened AMOC, and so the overshooting trajectories close to the critical rate see a larger increase in for example the pycnocline depth $D$ and $S_t$, similar to figure \ref{fig:laconstant}. These act as a stabilizing mechanism to the AMOC: a larger pycnocline depth implies a stronger AMOC while increased tropical salinity can salinify the northern box directly. Since these effects decrease for increasing rate, we expect the exceedance time $t_e$ to drop, and hence we obtain faster than logarithmic growth. 

\subsection{Comparison with a freshwater overshoot}
\begin{figure*}
    \centering
    \includegraphics[width = 39pc]{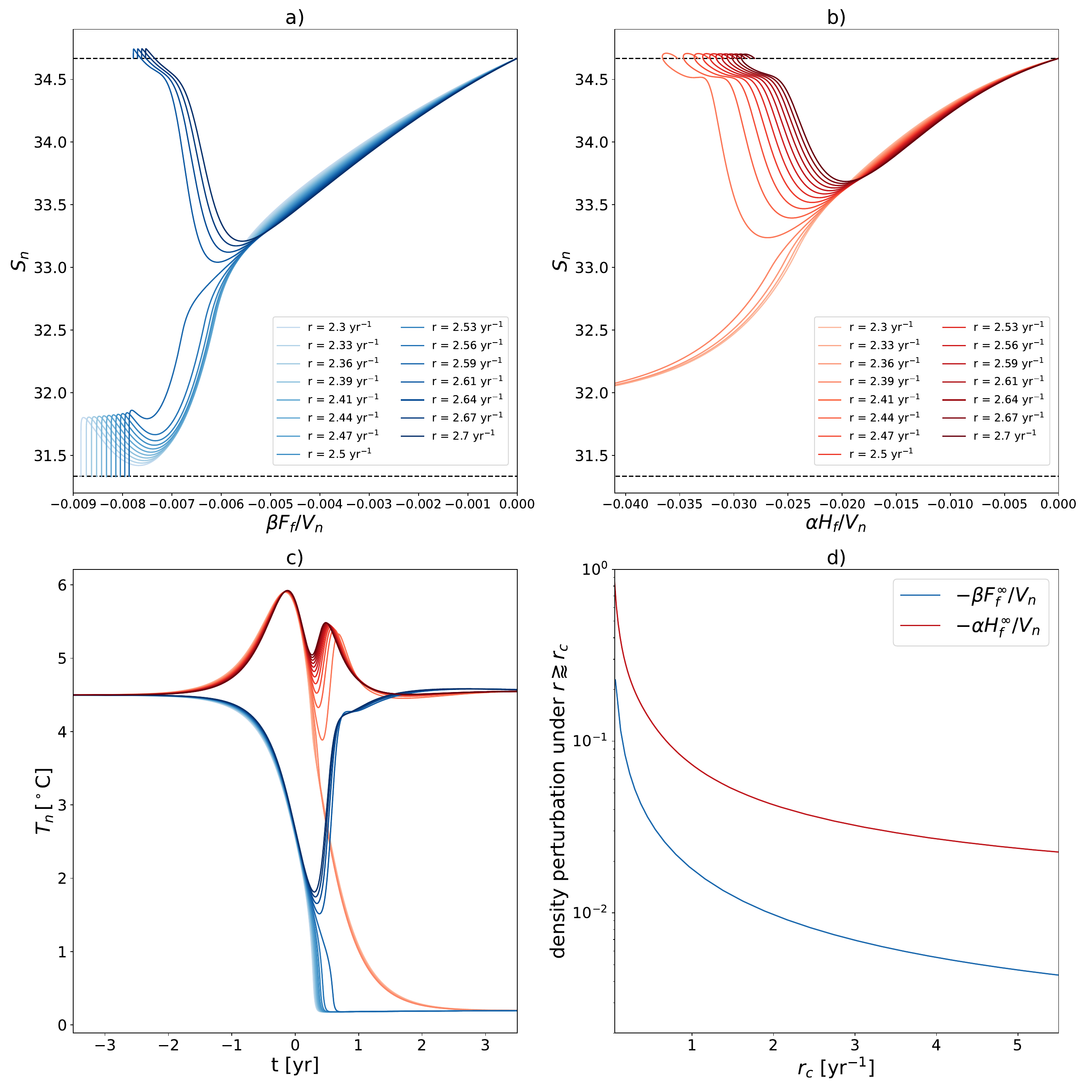}
    \caption{The northern salinity $S_n$ versus the non-dimensional density perturbation by a freshwater forcing $\beta F_f/V_n$ (a), and by a temperature forcing $\alpha H_f/V_n$ (b), where the exceedance amplitudes are $\varepsilon_a = 0.07$ Sv and $\theta_a = 2.1\,^\circ$C respectively, with the northern temperature $T_n$ along these same overshooting trajectories (c). The total non-dimensional density removed from the northern box under either temperature forcing $(-\alpha H_f^\infty/V_n)$ or freshwater forcing $(-\beta F_f^\infty/V_n)$ for overshooting trajectories that barely recovered ($r\gtrapprox r_c$) (d). The top and bottom dashed lines (a \& b) indicate the salinity of the ON and OFF state, respectively.}
    \label{fig:comp}
\end{figure*}

When comparing the dynamics of an overshoot in temperature versus one in freshwater forcing, we see that in both cases it is the salt-advection feedback that collapses the AMOC. In the former it is the weakening of the AMOC via thermal forcing that causes the initial drop in salinity $S_n$, while in the latter this is directly achieved via the freshwater flux. The salvaging factor of a higher rate of forcing in both cases is that it lowers $S_n$ quicker and hence the gyre can remove a higher portion of the freshwater perturbation, see figure \ref{fig:comp}a \& b. Here, initially for the same non-dimensional density perturbation ($\alpha H_f/V_n$ for temperature forcing and $\beta F_f/V_n$ for freshwater forcing) the trajectories under higher forcing rates have a lower $S_n$, which results in a stronger gyre transport. The main difference is in the ocean's northern temperature $T_n$: in a collapse under freshwater it drops as the AMOC's heat transport is reduced, whereas under temperature forcing it initially increases before dropping due to the lack of this transport, see figure \ref{fig:comp}c. The result is that under a temperature forcing the thermal feedback by reducing the AMOC is larger, since the temperature drop is bigger. This means that the AMOC is more resilient in a temperature overshoot than in a freshwater overshoot, which we can also deduce from figure \ref{fig:comp}d. Here the total removed non-dimensional density to the northern box is compared for trajectories that barely recover, i.e. the forcing rate is just higher than the critical rate $r_c$. It shows that the size of the density perturbation via thermal forcing is an order of magnitude larger than via freshwater forcing under the same forcing rates.

\section{Summary and Discussion}\label{sec:sum}
In this study we examined the physics of an AMOC overshoot addressing the critical rate that distinguishes a safe from an unsafe overshoot, and how this depends on the exceedance amplitude and the atmosphere-ocean coupling. Moreover, we have analyzed if it is crucial whether this forcing is a freshwater perturbation or a polar amplification of atmospheric temperature. A freshwater forcing that changes with a high enough rate does not collapse the AMOC, because the initial faster decrease in northern salinity results in a greater gyre transport for the same total amount of freshwater added. This gyre transport in turn removes this additional freshwater perturbation. The same mechanism also applies in an overshoot of the northern atmospheric temperature. Here the initial thermal forcing causes a decrease of the AMOC after which the main dynamics are salinity-driven. A higher rate of forcing causes a sharper AMOC decline which in turn lowers the northern salinity faster, and hence the gyre transport is again enhanced for the same total density perturbation. However, since in the temperature overshoot the drop in northern ocean temperature is larger, the stabilizing thermal feedback is boosted, and so the AMOC is more resilient to a temperature overshoot than to a freshwater overshoot under equal forcing rates. 

Regarding the AMOC's resilience to variations in the exceedance amplitude we find the same behaviour for both temperature and freshwater forcing. For low amplitudes the critical rate is low as well, resulting in a quasi-equilibrium overshoot. Here other state variables besides the ones directly affected by the forcing adjust as well. For larger amplitudes these adjustments can no longer occur as the exceedance time decreases resulting in a fast increase in critical rates. Eventually the critical rates of forcing are too large for the rest of the basin to adjust and the dynamics mostly occur in the northern box. The total amount of freshwater added or the total time of increased temperature determine whether the AMOC collapses or not. In this regime a stronger atmosphere-ocean coupling reduces the AMOC's resilience as it diminishes the effect of the thermal feedback during an AMOC decline. Conversely, close to the bifurcation point the stronger coupling actually enhances the resilience, since a larger coupling means that the AMOC is more thermally and less haline driven, and so less sensitive to freshwater forcing. Moreover, in this quasi-equilibrium setting the thermal feedback is not as effective since the drop in northern temperature is not as large (cf. figure \ref{fig:laconstant}d). 

In a recent study the dynamics of a freshwater overshoot were also analyzed in a global ocean model \citep{faure2025physical}. The problem set-up was different than here, but the mutual conclusion is that the northern salinity budget determines the recovery of the AMOC with indeed a crucial role for the lateral salt fluxes. Moreover, they propose the rate of change of the Atlantic salt content over time as an indicator for the outcome of the overshoot. This cannot be produced by our model as it only considers the Atlantic basin where salt conservation is assumed. In addition, a simple indicator for a safe overshoot does not seem feasible based on the presented results, since depending on the exceedance amplitude a safe overshoot could still see an almost collapsed AMOC with a freshened northern basin. 

The limitations of our study are not only the simplicity of our model, but also the set-up of the forcing. We assume a purely symmetric forcing in time, which is not necessarily realistic but makes a comparison with previous results \citep{ritchie2021overshooting} more straightforward. This choice in forcing also produces a rate parameter that quantifies the rate of change of any external forcing, which allows us to compare the temperature and freshwater forcing directly \citep{ritchie2023rate}. Lastly, our results are dependent on the choice of starting state, although the results should only change quantitatively as long as the starting state is in the bistable regime.

These findings can be applied in a wider context. As it is becoming more likely that we may have transgressed the $1.5\,^\circ$C goal of the Paris Agreement these recent years \citep{samset20242023}, an overshoot scenario becomes more likely. In particular as negative emission techniques \citep{chiquier2025integrated} and solar radiation management (SRM) scenarios \citep{irvine2016overview} become better understood. Several studies have already examined the AMOC's response to SRM, which also found a heightened relevance of the air-sea heat exchange in stabilizing the AMOC \citep{xie2022impacts}, and the absence of stabilizing mechanisms during a rapid cool-down of the atmospheric temperatures \citep{pfluger2024flawed}, which are in agreement with the physical mechanisms we found here. Moreover, the dynamical response of the AMOC to a freshwater overshoot \citep{faure2025physical} is qualitatively the same as to a temperature overshoot, and so these results can also be applied in the context of various emissions scenarios. Of course in an actual overshoot the forcing will be a combination of both freshwater and heat forcing.






\appendix[A]\label{app:A}
\appendixtitle{Model equations}
The box volumes $V_n$ and $V_s$ are fixed, while the other depend on the pycnocline depth $D$ according to 
\begin{align*}
    V_t &= A_tD\\
    V_{ts} &= \frac{1}{2}L_{xA}L_yD\\
    V_d &= V_0 - V_t - V_{ts} - V_n - V_s
\end{align*}
with the total volume $V_0$ of all 5 boxes constant. Here $A_t$, $L_{xA}$ and $L_y$ are the horizontal area of the Atlantic pycnocline i.e. of the tropical box, the zonal extant of the Atlantic Ocean at its southern end and the meridional extent of the frontal region of the Southern Ocean respectively. Note that the horizontal area of the south-tropical box follows as $A_{ts} = L_{xA}L_y$. The volume fluxes obey the following equations:
\begin{align*}
    q_n &= \eta \frac{\rho_n-\rho_{ts}}{\rho_0}D^2\\
    q_s &= q_{Ek} - q_e\\
    &= q_{Ek} - A_{GM}\frac{L_{xA}}{L_y}D\\
    q_u &= \kappa\frac{A_t}{D},
\end{align*}
where $\kappa$ and $A_{GM}$ fixed vertical and eddy diffusivity, and the equation of state is the commonly used:
\begin{align*}
    \rho_i = \rho_0(1-\alpha(T_i-T_0)+\beta(S_i-S_0)),
\end{align*}
so the northern downwelling becomes:
\begin{align*}
    q_n = \eta(\alpha(T_{ts}-T_n) + \beta(S_{n}-S_{ts}))D^2,
\end{align*}
where it is emphasized in \cite{cimatoribus2014meridional} that the northern downwelling should scale by a negative density difference between the southern tropical box \textit{ts} and the northern box ($\rho_{ts}<\rho_n$), and be non-existent otherwise ($\rho_{ts}\geq\rho_n$). Here $\eta$, $\alpha$ and $\beta$ are fixed coefficients. This results in the following transport equations:
\begin{align*}
    \der{(V_tS_t)}{t} &= 
    \underbrace{r_s(S_{ts}-S_t) + r_n(S_n-S_t)}_{\text{transport through gyres}} + \underbrace{2E_sS_0}_{\text{fresh water transport}}\\
    &+ \underbrace{q_s\big(\theta(q_s)S_{ts}+\theta(-q_s)S_t\big) + q_u S_d - \theta(q_n)q_nS_t}_{\text{transport through volume fluxes}}\\
    \der{(V_{ts}S_{ts})}{t} &= \underbrace{q_{Ek}S_s - q_eS_{ts}-q_s\big(\theta(q_s)S_{ts}+\theta(-q_s)S_t\big)}_{\text{transport through volume fluxes}}\\
    &\,\,+ \underbrace{r_s(S_{t}-S_{ts})}_{\text{transport through gyres}}\\
    \der{(V_nS_n)}{t} &= \underbrace{\theta(q_n)q_n(S_t-S_n)}_{\text{transport through volume fluxes}}\\
    &\,\,+ \underbrace{r_n(S_t-S_n)}_{\text{transport through gyres}} - \underbrace{(E_s+E_a)S_0}_{\text{fresh water transport}} \\
    \der{(V_sS_s)}{t} &= \underbrace{q_s\big(\theta(q_s)S_d+\theta(-q_s)S_s\big) +q_eS_{ts} - q_{Ek}S_s}_{\text{transport through volume fluxes}}\\
    &\,\,- \underbrace{(E_s-E_a)S_0}_{\text{fresh water transport}} \\
    \der{(V_dS_d)}{t} &= \underbrace{- q_s(\theta(q_s)S_d+\theta(-q_s)S_s)}_{\text{transport through volume fluxes}}\\
    &\,\,+\underbrace{\theta(q_n)q_nS_n -q_uS_d}_{\text{transport through volume fluxes}}
\end{align*}
with $S_0$ the average salinity of the Atlantic. Note that the total salinity content of the basin is conserved. Furthermore, the evolution of the pycnocline depth follows from volume conservation:
\begin{align*}
    V_r\der{D}{t} &= q_u + q_s - \theta(q_n)q_n\\
    &\text{ where }V_r = A_t + \frac{A_{ts}}{2},
\end{align*}
where $V_rD$ is then the total volume of the tropical boxes \textit{ts} and \textit{t}. Note that $\theta(x)$ is the Heaviside function, with $\theta(x) = 1$ for $x>0$ and $\theta(x) = 0$ for $x\leq 0$. Next, the evolution equations for the ocean's heat content are added:
\begin{align*}
    \der{(V_tT_t)}{t} &= \underbrace{q_s\big(\theta(q_s)T_{ts}+\theta(-q_s)T_t\big) + q_u T_d - \theta(q_n)q_nT_t}_{\text{transport through volume fluxes}}\\
    &\,\,+ \underbrace{r_s(T_{ts}-T_t) + r_n(T_n-T_t)}_{\text{transport through gyres}} - \underbrace{\lambda^a A_t(T_t - T_t^a )}_{\substack{\text{atmospheric heat}\\\text{ exchange}}}\\
    \der{(V_{ts}S_{ts})}{t} &= \underbrace{q_{Ek}T_s - q_eT_{ts}-q_s\big(\theta(q_s)T_{ts}+\theta(-q_s)T_t\big)}_{\text{transport through volume fluxes}}\\
    &\,\,+ \underbrace{r_s(T_{t}-T_{ts})}_{\text{transport through gyres}}  - \underbrace{\lambda^a A_{ts}(T_{ts} - T_{ts}^a)}_{\text{atmospheric heat exchange}}\\
    \der{(V_nT_n)}{t} &= \underbrace{\theta(q_n)q_n(T_t-T_n)}_{\text{transport through volume fluxes}} + \underbrace{r_n(T_t-T_n)}_{\text{transport through gyres}}\\
    &\,\,- \underbrace{\lambda^a A_n(T_n - T_n^a)}_{\text{atmospheric heat exchange}}\\
    \der{(V_sT_s)}{t} &= \underbrace{q_s\big(\theta(q_s)T_d+\theta(-q_s)T_s\big) +q_eT_{ts} - q_{Ek}T_s}_{\text{transport through volume fluxes}}\\
    &\,\,- \underbrace{\lambda^a A_s(T_s - T_s^a)}_{\text{atmospheric heat exchange}}\\
    \der{(V_dT_d)}{t} &= \underbrace{- q_s\big(\theta(q_s)T_d+\theta(-q_s)T_s\big)}_{\text{transport through volume fluxes}}\\
    &+\,\,\underbrace{\theta(q_n)q_nT_n - q_uT_d}_{\text{transport through volume fluxes}},
\end{align*}
where $A_{t}$, $A_{ts}$, $A_n$ and $A_s$ denote the surface areas of the surface boxes. Note that --unlike salinity-- the total heat content of the ocean is not conserved, as there is a net heat exchange with the atmosphere. 

\appendix[B]\label{app:B}
\appendixtitle{Parameter Values}
We non-dimensionalize the equations by scaling length, depth, time, salinity and temperature by $L$, $H$, $t_d$, $S_0$ and $T_0$ respectively. The parameter values are given in table \ref{tab:par1}. Throughout this work we will present results in dimensional units, while the actual computations are done non-dimensionally. The various values of the atmospheric temperatures and heat exchange constant are presented dimensionally in \ref{tab:par2}.

\begin{table*}
    \centering
    \begin{tabular}{llll}
    \toprule
    parameter & dimensional & non-dimensional & description\\
    \cmidrule{1-4}
     $L$ & $10^6$ m & 1 & length scale\\
     $H$ & $10^3$ m & 1 & depth scale\\
     $t_d$ & $1$ yr & 1 & time scale\\
     $S_0$ & $35$ psu & 1 & salinity scale\\
     $T_0$ & $1$ K & 1& temperature scale\\
     $V_0$ & $3\e{17}$ m$^3$ & 300 & volume Atlantic basin\\
     $V_n$ & $3\e{15}$ m$^3$ & 3 & volume box \textit{n}\\
     $V_s$ & $9\e{15}$ m$^3$ & 9 & volume box \textit{s}\\
     $A_t$ & $1\e{14}$ m$^2$ & 100 & surface area box \textit{t}\\
     $A_{ts}$ & $1\e{13}$ m$^2$ & 10 & surface area box \textit{ts}\\
     $A_n$ & $1\e{13}$ m$^2$ & 10 & surface area box \textit{n}\\
     $A_s$ & $3\e{13}$ m$^2$ & 30 & surface area box \textit{s}\\
     $L_{xA}$ & $10^7$ m & 10 & zonal extent South Atlantic boundary\\
     $L_y$ & $10^6$ m & 1 &meridional extent frontal Southern Ocean\\
     $A_{gm}$ & $1700$ m$^2$/s & 5.36112 & eddy diffusivity \\
     $q_{Ek}$ & $19.197$ Sv & 92.076 & northward Ekman transport\\
     $\kappa$ & $1\e{-5}$m$^2$/s & 0.031536 & vertical diffusivity\\
     $\eta$ & $3\e{4}$ m/s& 94608 & hydraulic constant\\
     $\alpha$ & $2\e{-4}$ $\degree $C$^{-1}$ & $2\e{-4}$ & thermal expansion coefficient\\
     $\beta$ & $8\e{-4}$ psu$^{-1}$ & $2.8\e{-2}$ & haline contraction coefficient\\
     $r_s$ & $10$ Sv & 31.536 & transport by southern subtropical gyre\\
     $r_n$ & $5$ Sv & 15.768 & transport by northern subtropical gyre\\
     $E_s$ & $0.17$ Sv & 0.536112 & symmetric freshwater flux\\
     $\lambda^a$ & $3.5\e{-6}$m/s & 11.0376 & heat exchange rate\\
     $T_{t}^a$ & $16.6774\degree$C & 16.6774 & atmospheric temperature \textit{t} box\\
     $T_{ts}^a$ & $16.6774\degree$C & 16.6774 & atmospheric temperature \textit{ts} box\\
     $T_{n}^a$ & $-2.0965\degree$C & -2.0965 & atmospheric temperature \textit{n} box\\
     $T_{s}^a$ & $-2.0965\degree$C & -2.0965 & atmospheric temperature \textit{s} box\\
    \hline
    \end{tabular}
    \caption{Used Parameter Values}
    \label{tab:par1}
\end{table*}

\begin{table*}
    \centering
    \begin{tabular}{lllll}
    \toprule
    $\lambda^a$[m/s] & $T_{t}^a$[$\degree$C] & $T_{ts}^a$[$\degree$C] & $T_{n}^a$[$\degree$C] & $T_{s}^a$[$\degree$C]\\
    \cmidrule{1-5}
    $3.0\e{-6}$ & 18.8099 & 18.8099 & -4.6504 & -4.6504\\
    $3.5\e{-6}$ & 16.6774 & 16.6774 & -2.0965 & -2.0965\\
    $4.0\e{-6}$ & 15.3801 & 15.3801 & -0.5762 & -0.5762\\
    $4.5\e{-6}$ & 14.5081 & 14.5081 & 0.4257& 0.4257\\
    $5.0\e{-6}$ & 13.8816 & 13.8816 & 1.1325 & 1.1325\\
    $5.5\e{-6}$ & 13.4096 & 13.4096 & 1.6561 & 1.6561\\
    $6.0\e{-6}$ & 13.0411 & 13.0411 & 2.0588 & 2.0588\\
    $6.5\e{-6}$ & 12.7454 & 12.7454 & 2.3774 & 2.3774\\
    $7.0\e{-6}$ & 12.5027 & 12.5027 & 2.6354 & 2.6354\\
    $7.5\e{-6}$ & 12.2999 & 12.2999 & 2.8485 & 2.8485\\
    $8.0\e{-6}$ & 12.1279 & 12.1279 & 3.0272 & 3.0272\\
    $8.5\e{-6}$ & 11.9801 & 11.9801 & 3.1792 & 3.1792\\
    $9.0\e{-6}$ & 11.8517 & 11.8517 & 3.3099 & 3.3099\\
    $9.5\e{-6}$ & 11.7391 & 11.7391 & 3.4234 & 3.4234\\
    $10.0\e{-6}$ & 11.6395 & 11.6395 & 3.5230 & 3.5230\\
    \hline
    \end{tabular}
    \caption{Heat Exchange Parameter Values}
    \label{tab:par2}
\end{table*}

%

\acknowledgments
J.S. and R.M.v.W. and H.A.D. are funded by the European Research Council through ERC-AdG project TAOC (project 101055096). T.G. acknowledges support from EPSRC projects EP/T011866/1 and
EP/V013319/1. For the purpose of open access, the authors have applied a
Creative Commons Attribution (CC BY) licence to any Author Accepted
Manuscript version arising from this submission.

%
%
\datastatement
The results can be readily reproduced using the described method and the stated parameter values. The code for the box model can be obtained via https://zenodo.org/records/11182090.


\bibliographystyle{ametsocV6}
\bibliography{references}

\end{document}